\documentstyle[preprint,eqsecnum,aps,psfig]{revtex}
\newcommand{\bfr}{{\bf r}}
\newcommand{\bfrp}{{\bf r'}}
\begin{document}
\bibliographystyle{unsrt}
%\draft
\title{Thomas-Fermi generalized approach for studying systems under pressure.} 
\author{E.Cappelluti\footnote{Corresponding author; e-mail:
 emmcapp@pil.phys.uniroma1.it}\\Dipartimento di Fisica,
Universita' "La Sapienza"\\
P.le Aldo Moro 5,
00185 Roma\\
Italy\\
L.Delle Site\\ 
Max-Planck-Institute for Polymer Research\\
Ackermannweg 10, D 55021 Mainz\\
Germany}
\date{\today}
\maketitle

\begin{abstract}
In a previous work one of the authors proposed a simple model
for studying systems under pressure based on the Thomas-Fermi (TF) model of 
single atom. In this work we intend to extend the previous work to 
more general Thomas-Fermi models where electronic exchange and
correlation are introduced. To do so, we first study numerically
the equation obtained by H.W.Lewis (TFDL) which introduces the effects
of exchange and correlation into the original TF equation; next the
procedure followed in the previous work is extended to the new
approach and a specific example is illustrated. Although one could
expect that no big differences were produced by the generalized TF
model, we show the qualitative as well as quantitative equivalence
with detailed numerical results. These results support the robustness
of our conclusions with regards to the model proposed in the previous
work and give the character of universality (i.e. to pass from one
atom to another, the quantities calculated must be simply scaled by
a numerical factor) to the properties of compressed systems shown in
this work. 
\end{abstract}
\newpage
\section{Introduction}
The statistical model of atom developed by E.Fermi \cite{fermi} and
known as Thomas-Fermi model, although based on a highly
simplified theoretical framework, has been proven surprisingly good in 
predicting basic properties of condensed matter \cite {fmt}. A
particular feature of such a model is the description of compressed
atoms, in this case the predicted properties were confirmed
experimentally \cite{exper}. In a preceding publication \cite{luigi}, 
we focused the attention on this particular aspect of the theory and
proposed a simple model for describing systems under pressure. The
central point of that work was the development of the concept of
``statistical ionization''. In simple terms, this is a universal
semianalytical function which enables one to describe, as a function
of the distance from the point-like nucleus, the balancing process
between the antibinding and binding contribution to the total energy
within the compressed atom. In spite of the approximations done and
extensively discussed, we underlined the utility of the proposed model 
as a tool for investigating at a basic level and at low computational
cost, properties of systems under pressure. However it was also
underlined that the properties of semianalycity and universality of
the function disappear when higher order of approximation are
introduced in the basic TF theory. In the light of what stated before, 
in this publication we intend to extend the treatment of the previous
work to more sophisticated models of the TF approach. That is to say,
to include the effects of exchange and correlation into the original
TF model and obtain in this framework a ``generalized statistical
ionization function''. The extension of the original TF model is due
to P.A.M.Dirac (TFD) \cite{dirac} 
for the exchange while  for the part relative to the correlation,
several approaches has been proposed; in this work the one proposed by 
H.Lewis (TFDL) \cite{lewis} has been chosen because it is simple and
appropriate for the compressed case, since the treatment of the
electrons becomes exact in high density limit (see appendix). The
paper is organized as follows; first we obtain the TF, TFD and TFDL
equation within a single generalized approach, then we show numerical
solutions for different atomic numbers in the neutral uncompressed
case, next we illustrate numerical results for the ``generalized
statistical ionization function'' in case of $Z=50$ ($Z$ atomic
number). Finally the equation of state of a compressed system is
calculated. Comments on the results obtained as well as on 
advantages and limitations of this model,conclude the work.

\section{A statistical approach: the Thomas-Fermi-Dirac-Lewis model}
In this section we derive a unique general form for the TF, TFD and
TFDL equation.
In a semiclassical approximation the local electronic
density of states can be defined as (see for example \cite{landau}):
\begin{equation}
dn[p(\bfr),(\bfr)]=\frac{8\pi}{h^3}p^2(\bfr) dp(\bfr).
\end{equation}
The local electron density is therefore given by:
\begin{equation}
n(\bfr)=\int_0^{p_F(\bfr)}\frac{8\pi}{h^3}p^2(\bfr) dp(\bfr)
=
\frac{8\pi}{3}\frac{p_{\rm F}^3(\bfr)}{h^3},
\label{nr}
\end{equation}
which determines the local Fermi vector $p_{\rm F}(\bfr)$.

In the spirit of the (generalized) Thomas-Fermi approach we can
express the one-particle energy $E[p(\bfr),\bfr]$ through an effective
Hamiltonian $H_{\rm el}[p(\bfr)]$ of free (interacting) electrons
plus an electrostatic field $V(\bfr)$ arising from the direct electron-nucleus
and electron-electron interaction (we must remind that the nucleus is 
considered a positive point-like charge):
\begin{equation}
E[p(\bfr),\bfr] = H_{\rm el}[p(\bfr)]-eV(\bfr),
\end{equation}
For a system at the equilibrium the chemical potential 
$\mu(\bfr)$, defined by the maximum energy
of $\max\{E[p(\bfr),(\bfr)]\}=E[p_{\rm F}(\bfr),\bfr]$,
has to be a constant $\mu$ independent of $\bfr$.
Putting for convenience $\mu = - e V_0$
we obtain therefore:
\begin{equation}
H_{\rm el}[p_{\rm F}(\bfr)]= e[V(\bfr) - V_0].
\end{equation}

The effective field $V(\bfr)$ will be determined self-consistently
via electrostatic considerations by using the Poisson's equation:
\begin{equation}
\nabla^2[V(\bfr)-V_0]=
\frac{1}{r}\frac{d^2}{dr^2}\left\{r\left[V(r)-V_0\right]\right\}
=e\frac{n(r)}{\epsilon_{0}},
\label{poissonv}
\end{equation}
where $r=|\bfr|$.
Eq. (\ref{poissonv}) can be quite simplified by introducing the variables
$\phi(r)$ and $v(r)$ defined by:
\begin{equation}
\frac{Ze}{4 \pi \epsilon_{0} r} \phi(r) = V(r)-V_0,
\end{equation}
\begin{equation}
v^2(r)=(2Ze^2 m)^{-1} (4 \pi \epsilon_{0}) r p_{\rm F}^2(r).
\end{equation}
Eq. (\ref{poissonv}) becomes thus:
\begin{equation}
\phi''(r) = \frac{1}{Z \epsilon_{0}} \left(\frac{8\pi}{3h^3}\right)
(2Z E^2 m)^{3/2}(4 \pi \epsilon_{0})^{-1/2} v^3(r) r^{-1/2},
\end{equation}
and finally, by scaling $r \rightarrow \beta a_0 x$, we have:
\begin{equation}
\phi''(x)=v^3(x) x^{-1/2},
\label{firsteq}
\end{equation}
this is the ``generalized TF equation'', where
\begin{equation}
a_0=\frac{4 \pi \epsilon_{0} h^2}{4\pi^2 m e^2},
\end{equation}
is the atomic Bohr radius and
\begin{equation}
\beta=\left(\frac{9\pi^2}{128 Z}\right)^{1/3}.
\end{equation}

A closed system of differential equations 
is now obtained by expressing back the variable $\phi(x)$ as function
of $v(x)$. To this aim the explicit expression
of the electronic Hamiltonian $H_{\rm el}[p(r)]$ is needed.
Of course this task is in principle quite hard since the solution 
of a many-body system is necessary, and some kind of approximation
is required. The original TF model represents 
the simplest approximation, i.e. the Hamiltonian $H[p(r)]$
is approximated with the only kinetic term:
\begin{equation}
H^{\rm TF}_{\rm el}[p(r)]=\frac{p^2(r)}{2m}.
\label{hptf}
\end{equation}
The total Hamiltonian contains therefore only kinetic and
the direct Coulomb electron-electron interaction, totally neglecting
any quantum contribution.
By using again the reduced variables $\phi(x)$ and $v(x)$,
we can write Eq. (\ref{hptf})  in the compact form:
\begin{equation}
\phi(x)=v^2(x),
\label{tfphi}
\end{equation}
which, together with Eq. (\ref{firsteq}), defines the 
statistical solution of the Thomas-Fermi model.

AS stated before higher order of approximation have been introduced in
literature to include quantum exchange and correlation contribution
to Eq. (\ref{hptf}). Most famous is the so called Thomas-Fermi-Dirac (TFD)
model which explicitly takes into account the exchange energy (see
\cite{dirac,fmt}).
The inclusion of the correlation term has been a more debated issue,
and different approaches have been proposed. A particular suitable
and simple one is the interpolation formula given by H.W. Lewis,
which becomes exact in the both the high-density and
low-density limits \cite{lewis}. We referred and will refer to it
as the TFDL model; in this case the Hamiltonian contains
kinetic, exchange and correlation terms:
\begin{equation}
H^{\rm TFDL}_{\rm el}[p(r)]=
\frac{p^2(r)}{2m}
-\frac{e^2p(r)}{4 \pi \epsilon_{0}\pi \hbar}
-\frac{me^4(1-\ln2)}{(4 \pi \epsilon_{0} \pi \hbar)^2}
\ln\left[1+
\frac{(0.89 \alpha \pi -1)\pi p(r) a_0}{(1-\ln 2)\hbar}\right],
\label{eqlewis}
\end{equation}
where
\begin{equation}
\alpha=\left(\frac{4}{9\pi}\right)^{1/3}.
\end{equation}
Introducing the variables $\phi(x)$ and $v(x)$, Eq. (\ref{eqlewis})
reads:
\begin{equation}
\phi(x)=v^2(x)-av(x)x^{1/2}-\rho a^2 x \ln\left[\sigma a+v(x)x^{-1/2}\right],
\label{tfdlphi}
\end{equation}
where
\begin{equation}
a=\left(\frac{2\beta}{\pi^2 Z}\right)^{1/2},
\end{equation}
\begin{equation}
\sigma=\frac{1-\ln 2}{2(0.89 \alpha \pi -1)},
\end{equation}
and
\begin{equation}
\rho=\frac{1}{2}(1-\ln 2).
\end{equation}

Eq. (\ref{tfdlphi}) can be considered a generalization of the TF
model [Eq. (\ref{tfphi})] to include exchange and correlation energies.
We find Lewis's formula particularly appealing since we are
mainly interested in the high-density regime of compressed atoms where
the interpolation formula provided by Eq. (\ref{eqlewis})
is expected to work quite well. In spite of the simplicity of the
model, to our knowledge, no explicit numerical or analytical treatment 
of this equation has been published, although several authors
\cite{barnes1,barnes2,salpeter,ebina} cite this approach in their
work. The main reason why this happens is due to the fact that since the
TFD model does not produce evident changes with respect to the simple TF
solutions, the TFDL is expected to not be much different.  

However in our paper we derive for the first time explicit numerical 
solutions of
the TFDL model and 
we use it as basic starting point of our analysis;
the results will be compared with the simple TF model.
Although we do not expect large differences with respect to the TF, we 
carry on this study to confirm a qualitative and quantitative
equivalence between TF and TFDL and later on the consequent advantages 
of such a conclusion will be underlined.

\section{Numerical study of TFDL equation and ``generalized ionization 
  function''}
In this section we illustrate some TF,TFD and TFDL solutions for the
neutral uncompressed case, the trend shown for this case is preserved 
for the compressed case. Then we show within a generalized TF approach,
the procedure to obtain the ``statistical ionization function'' via
the total energy as a function of the distance from the nucleus. 
Eqs. (\ref{firsteq}) and (\ref{tfdlphi}), or
Eqs. (\ref{firsteq}) and (\ref{tfphi}), represent two sets of coupled
differential equations which can be solved to obtain $\phi(x)$ or,
equivalently, $v(x)$, and as a consequence
the local electron density $n(r)$. The boundary
conditions required to determine a unique solution
are provided by the asymptotic limit:
\begin{equation}
\lim_{r \rightarrow 0} \phi(x)=1,
\end{equation}
and by fixing the radius of
a neutral atom:
\begin{equation}
\phi'(x_0)=x_0 \phi(x_0).
\end{equation}

Although it has been widely pointed out in literature,
it is important to remind that the inclusion 
of exchange and correlation terms breaks down the universality
of the solution. Namely, it is not possible to obtain local
electron density $n(r)$ for a given atomic number $Z$ just by a proper scaling
of the physical quantities (length, energy etc \ldots).
In Fig. \ref{fi-x}
we show the radial distribution of $\phi(x)$ for a neutral
uncompressed atom for the Thomas-Fermi model
and for the Thomas-Fermi-Dirac-Lewis model 
with statistically relevant values of atomic number:
$Z=10,50,80$. Only a slight discrepancy is found as result of the inclusion
of the exchange and correlation terms,
pointing out the almost quantitative accuracy of the TF model.
As expected, such a discrepancy vanishes as the atomic number $Z$
is increased. This result show that TF and TFDL are quantitatively as
well as qualitatively equivalent (the trend as stated before holds in
the compressed case as well).

By the knowledge of the solution $\phi(x)$ we can now evaluate
the semiclassical expression of the local density of energy
$E(r)$ which is related to the one electron energy $E[p(r),r]$ through the
relation $E(r)= \int dn[p(r),(r)] E[p(r),r]$.
We can identify three different contributions:
\begin{equation}
E(r)=E_{\rm kin}(r)+E_{\rm e-i}(r)+E_{\rm e-e}(r),
\end{equation}
where
\begin{equation}
E_{\rm kin}(r)=\int dn[p(r),(r)] \frac{p^2(r)}{2m},
\end{equation}
\begin{equation}
E_{\rm e-n}(r)=-\frac{Ze^2}{4\pi \epsilon_{0} r}n(r),
\end{equation}
\begin{equation}
E_{\rm e-e}(r)=\frac{1}{2}\frac{e^2}{4\pi \epsilon_{0}}n(r)
\int_{|\bfrp|<r_0}  d\bfrp 
\frac{n(\bfrp)}{|\bfr-\bfrp|},
\end{equation}
where of course
\begin{equation}
n(r)=\int dn[p(r),(r)].
\end{equation}

A particularly meaningful quantity is also the integrated
energy defined $E(R)$ by:
\begin{equation}
E(R)=\int_0^R 4\pi r^2 dr E(r),
\end{equation}
where $R$ is a generic distance $R \le r_0$.
$E(R)$ represents therefore the total energy contained in the
atomic volume enclosed within the distance $R$ from the nucleus.

Just as $E(r)$, we can think $E(R)$ as composed by a kinetic, 
a Coulomb electron-nucleus and a Coulomb electron-electron term.
After some straightforward manipulations (see also \cite{teulosky}) 
, by using once more
the reduced variables, we can write each of these contributions as:
\begin{equation}
E_{\rm kin}(X)= \frac{2^{10/3}(3\pi)^{1/3}}{5\pi} Z^{7/3}
\left[\frac{me^4}{2(4 \pi \epsilon_{0} \hbar)^2}\right]
\int_0^X dx v^5(x) x^{-1/2},
\end{equation}
\begin{equation}
E_{\rm e-n}(X)=-\frac{2^{10/3}(3\pi)^{1/3}}{3\pi} Z^{7/3}
\left[\frac{me^4}{2(4 \pi \epsilon_{0} \hbar)^2}\right]
\int_0^X dx v^3(x) x^{-1/2},
\end{equation}
\begin{equation}
E_{\rm e-e}(X)=\frac{1}{2}\frac{2^{10/3}(3\pi)^{1/3}}{3\pi}
Z^{7/3}
\left[\frac{me^4}{2(4 \pi \epsilon_{0} \hbar)^2}\right]
\int_0^X dx v^3(x) x^{1/2}
\left[\frac{1}{x}\int_0^x v^3(x') {x'}^{1/2}
+ \int_0^{x_0} v^3(x') {x'}^{-1/2}
\right],
\end{equation}
where $X$ is the scaled variable related to $R$ through the relation
$R=\beta a_0 X$ and where
we made use of the relation:
\begin{equation}
\int_{|\bfr|<R}  \int_{|\bfr|<r_0} 
\frac{f(\bfr) f(\bfrp)}{|\bfr-\bfrp|} d\bfr d\bfrp
=
\int_0^R dr 4\pi r^2 f(r) 
\left[\frac{1}{r} \int_0^r dr' 4\pi {r'}^2 f(r') 
+ \int_0^{r_0} dr' 4\pi r' f(r')
\right],
\end{equation}
valid for an isotropic $f(\bfr)$.

As it was done in \cite{luigi}, we consider $R$ a variable distance
within the atom, and $E(R)$, as stated before, is the total energy in
classical terms, of the sphere of radius $R$ inside the spherical
atoms. In simple terms, we do divide the atom into infinitesimally
thin concentric shells; the energy at $R$ is the sum of the
contributions of all of the shells inside $R$. As a consequence the
distance at which $E(R)$ has its minimum can be interpreted as the
distance at which the binding (nucleus-electrons) and antibinding
(electron-electron and kinetic energy) contributions to the total
energy are in exact balance. This allows us to define a sort of
electron ionization criterion, 
where the term ``ionization'' stands for electrons with
zero or positive energy; thus we can address to $E(X)$ as the
``generalized statistical ionization function''. The exact meaning of
''statistical ionization'' with the related limitations has 
been extensively discussed in our previous work, for such a reason we
do not spend more discussion about it and focus the attention onto the 
numerical results.

The number of ionized electron is therefore given by
\begin{equation}
N_{\rm ion} = \int_{R \le |\bfr| \le r_0}
n(\bfr) d\bfr,
\end{equation}
which, written in reduced variables, reads:
\begin{equation}
N_{\rm ion} = Z \int_{x_{\rm core}}^{x_0} dx v^3(x) x^{1/2}.
\label{nions}
\end{equation}
Differently from the procedure adopted in our previous work, in this
case $E(X)$ can be studied only numerically, since it is not possible
to solve analytically the integrals and so obtain a semianalytical
expression. Moreover the dependence of $\phi(x)$ on $Z$ does not allow 
to obtain a universal function where to pass from one atom to another 
is possible by simply scaling in $Z$. Next we show results obtained by 
studying numerically the ``generalized statistical ionization
function'' for the case $Z=50$; however since we did find quantitative 
agreement between TF and TFDL, the results shown are valid for any
atomic number by opportunely scaling them.

\section{Results}
In this section we show results obtained for the ``generalized
statistical ionization function'' for the case $Z=50$. As it was
already underlined, we expect that the results are quantitatively and
qualitatively equivalent in case TF or TFDL model is used, since the
solution of the TF and TFDL equation do not differ, as it was shown in 
the previous section. This means that the ''generalized ionization
function'' is the numerical equivalent of the semianalytical and
universal ''statistical ionization function'' of our previous work;
thus the results (and the equation of state) 
shown here for $Z=50$, can be easily generalized to
any $Z$ by simply scaling. Indeed the function $E(X)$ does not show 
a different behavior when TF or TFDL are used; in figure
\ref{Etot_s-s} the plots obtained using TF and using TFDL coincide.
As it is possible to see in figure
\ref{Etot_s-s} a minimum is always obtained for any compression, and
in figure \ref{Etot_s=n_x} a particular compression has been chosen 
and the determination on the number of ionized electrons is
pictorially illustrated.

Accordingly to the procedure of our previous work, at this point we
can therefore model the compressed atom
as a core with radius $R_{\rm core}$ plus a number of ionized
electrons $N_{\rm ion}$ which in principle is free and is spread
over all the atomic volume. The pressure is therefore determined
by the density of `     `free'' (ionized) electrons as 
we proposed in the previous work (again limitations are extensively
discussed there):
\begin{equation}
P= \frac{(3\pi^2)^{2/3}}{5}\frac{\hbar^2}{m} 
\left(\frac{N_{\rm ion}}{V_{\rm atom}}\right),
\end{equation}
where $N_{\rm ion}$ is given by Eq. (\ref{nions})
and the atomic volume is simply
$V_{\rm atom}=r_0^3$.
First we study the ionization as a function of the compression for the 
single atom (see Fig.\ref{nion-xcore-x0}) and then we extend the
procedure to a macroscopic level so that the 
resulting equation of state is shown in Fig. \ref{P-Vf},
where the pressure in units of Rydberg over the Bohr atomic volume
$V_{\rm Bohr} = a_0^3$
is plotted as function of the volume.
\section{Discussion and Conclusion}
We generalized the concept of ``statistical ionization function'' for
electrons in a compressed atom, obtained in a previous work, within
the TF approach. The generalization was developed by extending the
previous analysis to more sophisticated TF models where exchange and
correlation are considered. We found that there is no qualitative and
quantitative difference between the original and the generalized
approach. Although one could have expected this result, we explicitly 
proved such an equivalence, and for the first time, solutions of the
TFDL equation where shown. The important conclusion of this work
concerns the fact that because of the equivalence shown, the results 
reported for the particular case of $Z=50$ are valid for any atomic
number, provided that a rescaling $\frac{Z^{*}}{Z=50}$, where $Z^{*}$
is the atomic number wished, is applied; above all the results
legitimate the use of the semianalytical and universal ionization
function obtained in our previous work with the evident
advantages of the low computational cost and its extreme simplicity
and immediacy. Moreover the fact that Lewis' formula is exact for the
high density limit, means that in our case the effects of correlation 
were well described, as a consequence one can conclude that such
effects are not relevant for describing atoms under pressure, at least 
in first approximation; this
result it is not obvious in an {\it a priori} analysis. As it was
discussed in the previous work, the equation of state is a
simplification and as a consequence far from being rigorous; for
instance an open problem of our model is the distribution of the
``ionized'' electrons which in the present work is considered simply
uniform. However, due to its simplicity and feasibility, the model can 
be applied for a basic study of compressed matter not only via the
determination of the equation of state, but also as a basis for
developing more efficient analysis within more sophisticated
theories. The example shown in our previous publication, was the
determination of the ``ionized'' electrons for a certain compression
and the consequent description, in an {\it ab initio} method, of the
ionized electrons via plane-waves wavefunctions while the other
electrons can be represented as a core or as localized orbitals; this
would speed up the convergence for self-consistent calculations of
compressed matter. Here we can say more, as it is well known, the {\it 
  ab initio} calculations are based on the concept of pseudopotential; 
only the valence electrons are explicitly taken into account while the 
rest are placed in a core described by an opportune potential which 
interacts with the valence electrons. Valence electrons and core are
known for the uncompressed atoms but one may ask what happens in case
the system is under high pressure, in this case the valence electrons
and the core should be redefined according to the degree of
compression. In this case our model, at basic level' would be very
helpful for estimating the number of ionized (valence) electrons and
define what is the core. The evident advantage of such a procedure
stays in the simplicity of such an estimate and in the very low
computational cost. In conclusion we think that the study performed in 
this work furnishes important information and tools for a
computational inexpensive and well founded basic analysis of
compressed matter.
\section{Appendix}
\subsection{Correlation Energy within Lewis' approach}
We briefly illustrate the procedure followed by lewis to develop
a suitable formula for the electron correlation, for more details see
\cite{lewis} and references therein.
Consider an electronic Fermi gas whose maximum momentum is the Fermi
momentum $p_{F}$ and density $n=\frac{p_{F}^{3}}{3\pi^{2}\hbar^{3}}$;
The correlation energy at high density can be calculated via the
Gell-Mann's scheme and leads to the following expression 
\begin{equation}
E_{corr}=-(me^{4}/\pi^{2}\hbar^{2})(1-ln2)ln(p_{F})+const
\label{formula1}
\end{equation}
$m$ and $e$ are respectively the electron mass and the electron charge.
This expression must be modified in such a way that its validity could 
be reasonably extended to any density and in particular must be exact
at low density. At this point Lewis invokes Wigner procedure for the
calculation of electron correlation for a dilute gas;
i.e. one simply needs to note that a very dilute electron gas in the
ground state crystallize into a body-centered cubic lattice and at
this point the correlation energy can be determined exactly via a
Madelung type technique. The expression for low density obtained is
\begin{equation}
E_{corr}=-(0.89\alpha\pi-1)e^{2}p_{F}/\pi\hbar
\label{formula2}
\end{equation}
where $\alpha=(4/9\pi)^{1/3}$. Finally the procedure leads to what
Lewis defines as ``a suitable interpolation formula'' for the
correlation energy valid for any density:
\begin{equation}
E_{corr}\approx
-\frac{me^{4}(1-ln2)}{\pi^{2}\hbar^{2}}ln\left[1+
\frac{(0.89\alpha\pi-1)p_{F}\pi a_{0}}{(1-ln2)\hbar}\right] 
\label{formula3}
\end{equation}
where $a_{0}=\hbar^{2}/m e^{2}$. This is the expression used by Lewis
to incorporate the correlation effects into the TFD model.

\begin{figure}
%\centerline{\psfig{figure=fi-x.eps,width=9cm}}
\caption{Radial distribution of $\phi(x)$ as function of $x$
for a neutral uncompressed atom. Dashed line represents the Thomas-Fermi
model, solid lines the Thomas-Fermi-Dirac-Lewis one with atomic
number (from lower to upper line) $Z=10,50,80$. The $Z=50$ and $Z=80$ cases
are strongly overlapping and are hardy distinguishable in the plot.}
\label{fi-x}
\end{figure}

\begin{figure}
%\centerline{\psfig{figure=Etot_s-s.eps,width=9cm}}
\caption{Integrated energy $E(R)$ as function of $R$ for the
Thomas-Fermi-Dirac-Lewis model ($Z=50$) and different degrees of compression
parameterized by the atomic radius (from lower to upper line):
$x_0=1.5,1.4,1.3,\ldots,0.9,0.8$. $E(R)$ always presents a minimum
at a  certain $R$.}
\label{Etot_s-s}
\end{figure}

\begin{figure}
%\centerline{\psfig{figure=Etot_s=n_x.eps,width=9cm}}
\caption{Top panel: particular of Fig. 2 for $x_0=1.2$.
The minimum of $E(R)$ determines the ``core'' radius $R_{\rm core}$.
Bottom panel: graphical representation of the number
of ionized electrons $N_{ion}$ corresponding to the
above case. $N_{ion}$ is determined
by integrating the electron density from the ``core'' radius
to the atomic boundary.}
\label{Etot_s=n_x}
\end{figure}

\begin{figure}
%\centerline{\psfig{figure=nion-xcore-x0.eps,width=9cm}}
\caption{Number of ionized electrons (solid line, left scale)
and core radius $x_{\rm core}$ (dashed line, right scale)
as function of the atomic radius
$x_0$.}
\label{nion-xcore-x0}
\end{figure}
\begin{figure}
%\centerline{\psfig{figure=P-V.eps,width=9cm}}
\caption{Pressure {\em vs.} volume dependence for the statistical ionization
model.}
\label{P-Vf}
\end{figure}
\clearpage
\renewcommand{\thepage}{fig\arabic {page}}
\setcounter{page}{1}
\begin{figure}
\centerline{\psfig{figure=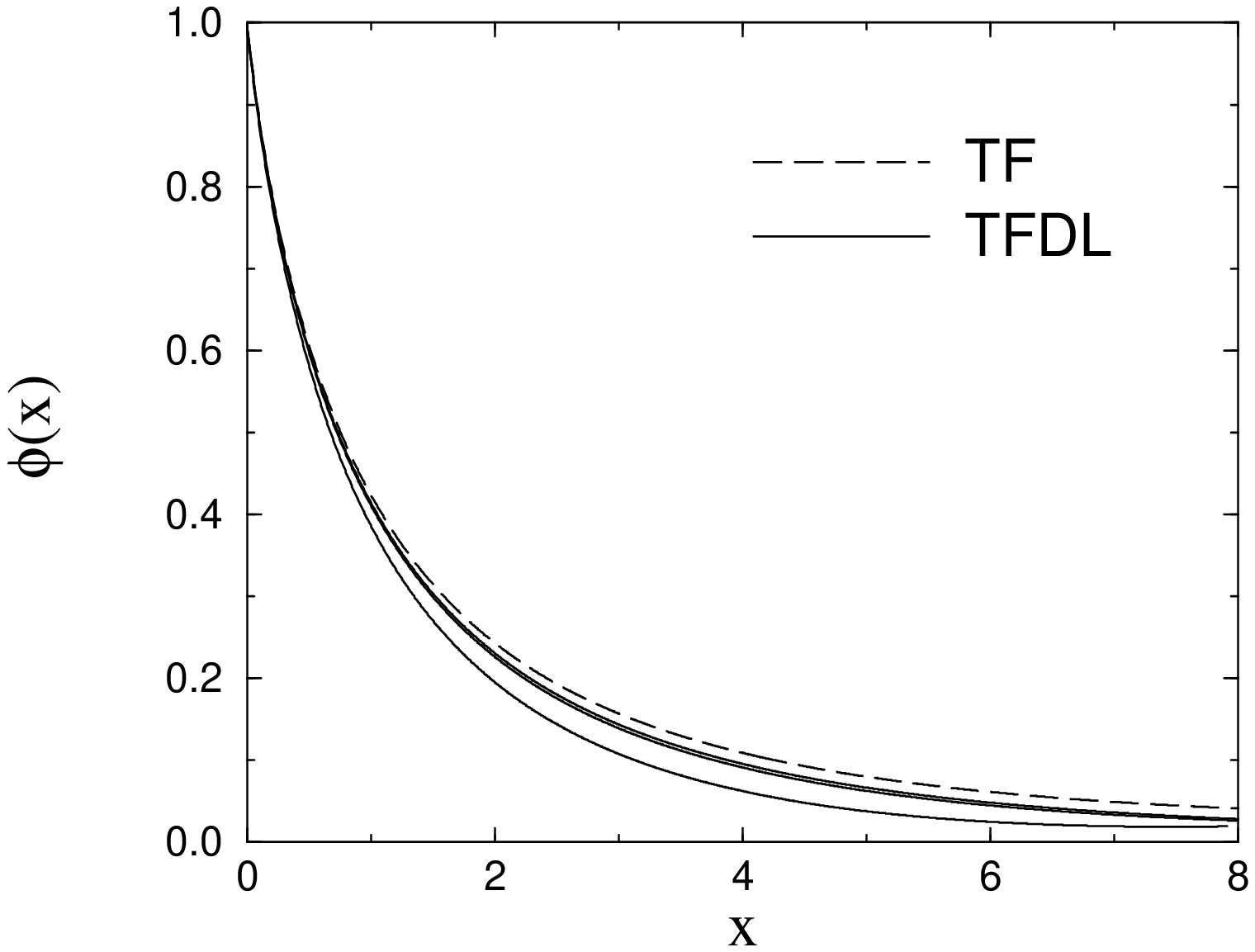}}
\end{figure}
\newpage
\begin{figure}
\centerline{\psfig{figure=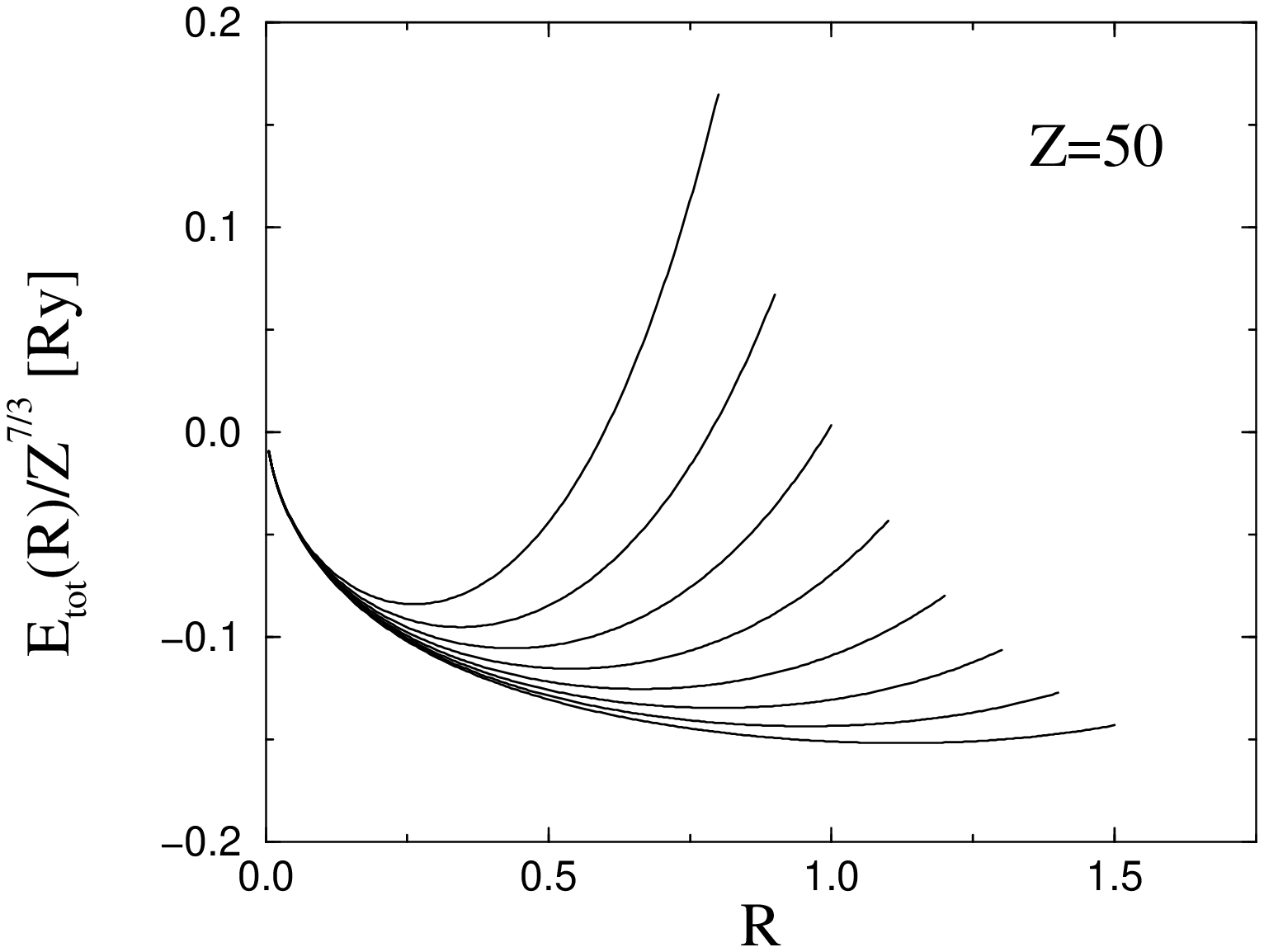}}
\end{figure}
\newpage
\begin{figure}
\centerline{\psfig{figure=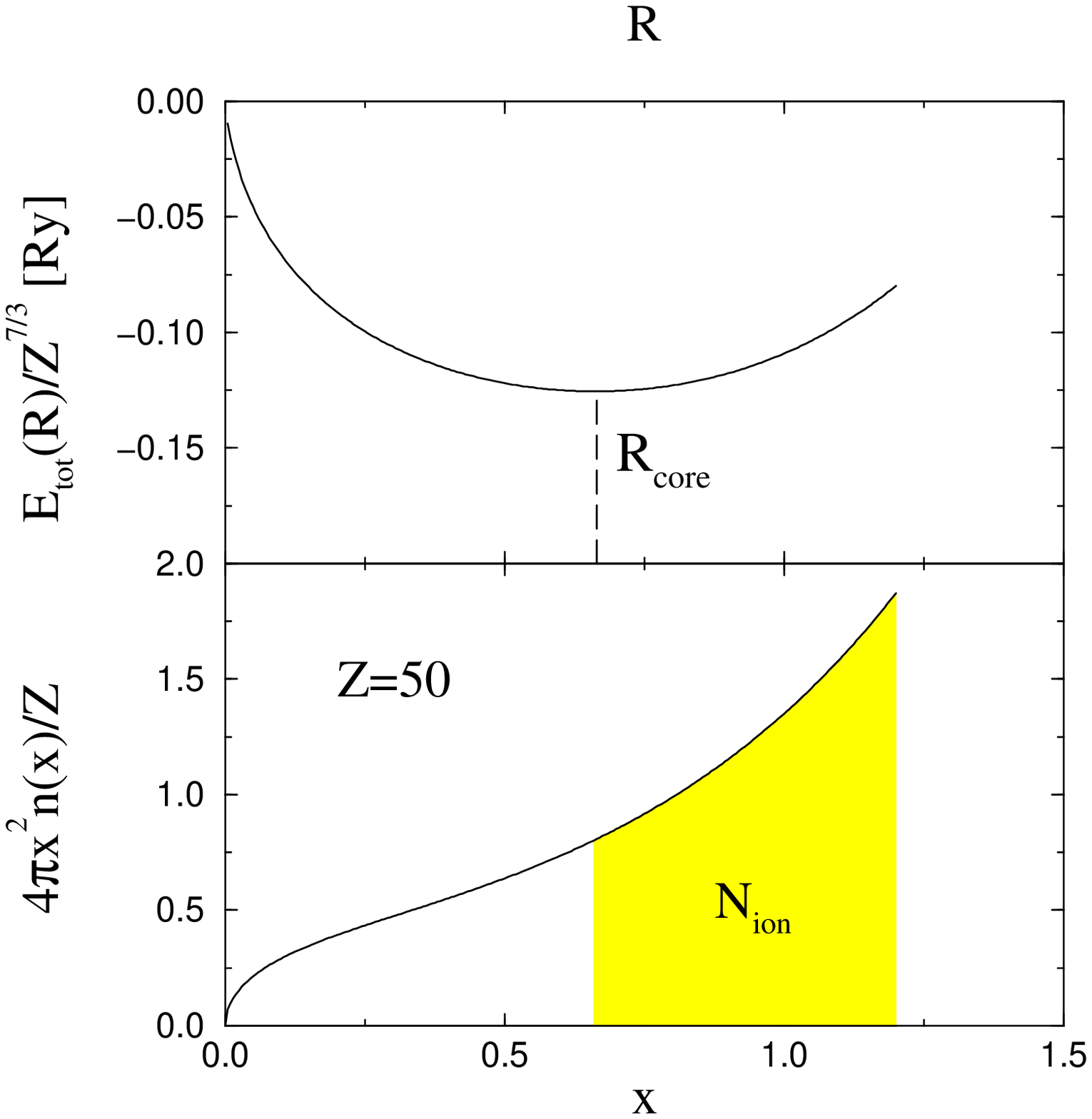}}
\end{figure}
\newpage
\begin{figure}
\centerline{\psfig{figure=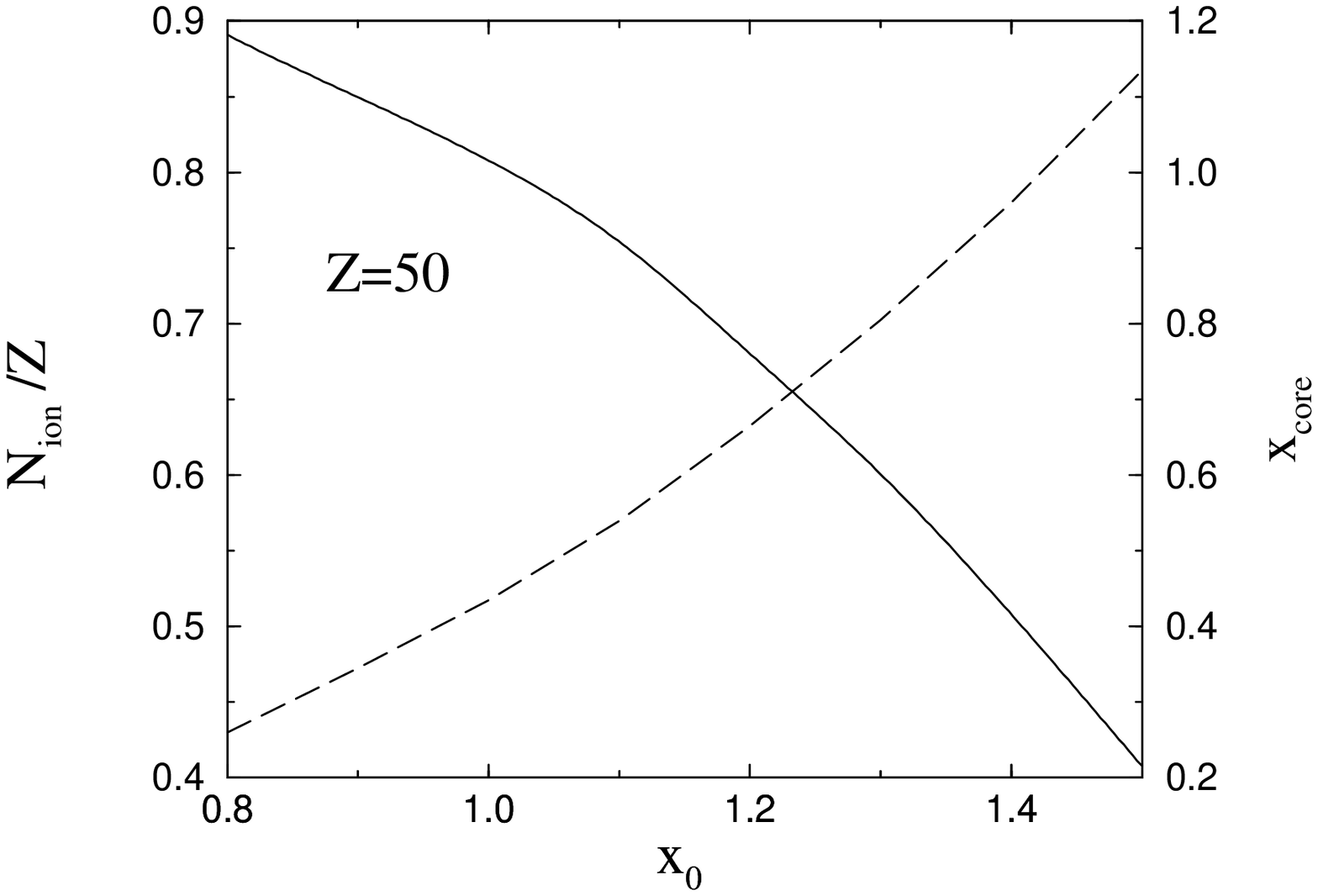}}
\end{figure}
\newpage
\begin{figure}
\centerline{\psfig{figure=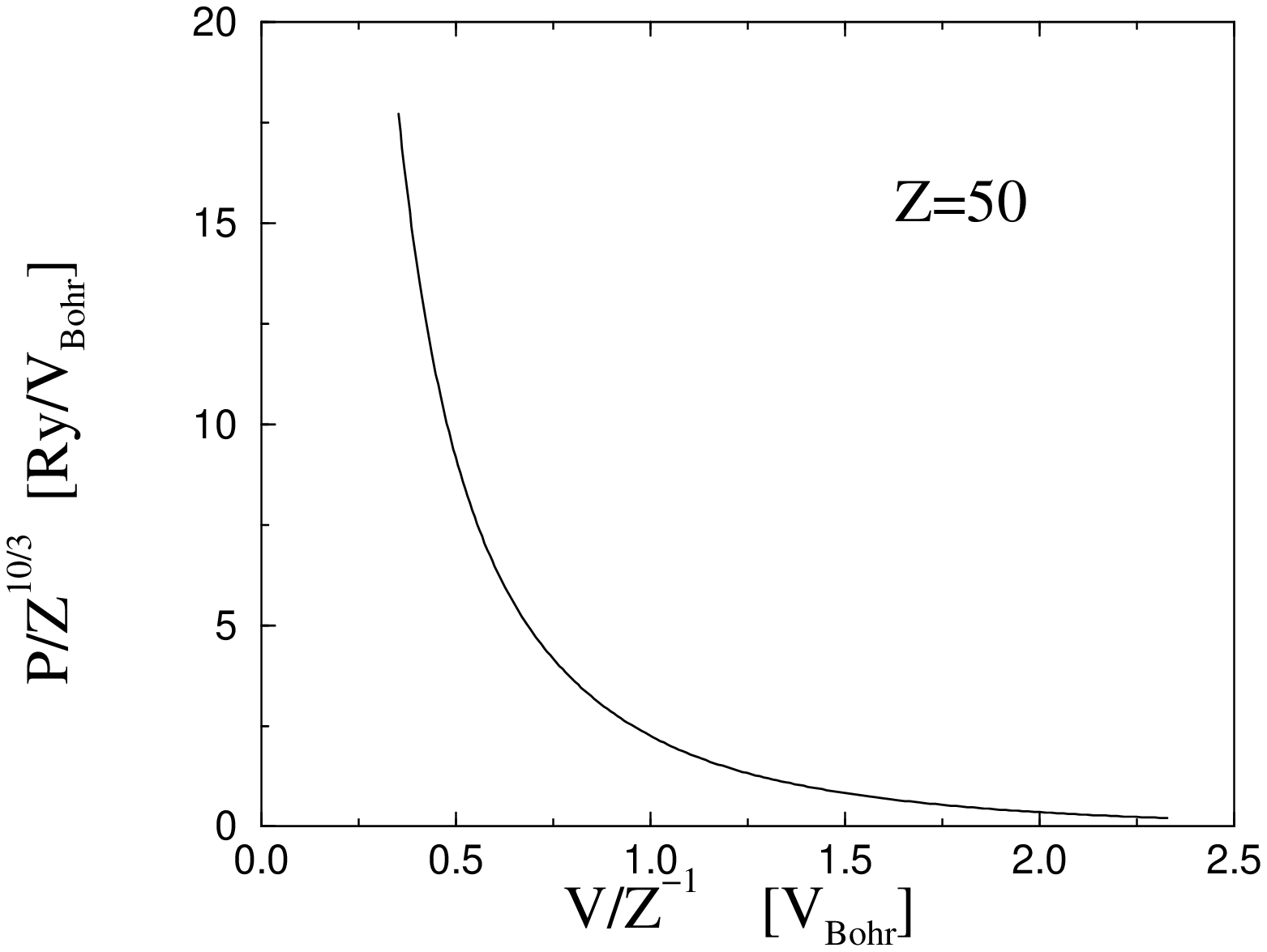}}
\end{figure}

\end{document}